\documentstyle[preprint,aps]{revtex}
\begin{document}
\title{
Long-Time Fluctuations in a Dynamical Model \\
of Stock Market Indices}
\author{
Ofer Biham$^1$,
Zhi-Feng Huang$^2$,
Ofer Malcai$^1$
and
Sorin Solomon$^1$}
\address{$^1$Racah Institute of Physics,
The Hebrew University,
Jerusalem 91904,Israel}
\address{$^2$Department of Physics, University of Toronto,
Toronto, Ontario M5S 1A7, Canada}

\maketitle
\newpage

\begin{abstract}

Financial time series typically exhibit strong 
fluctuations that cannot be described by a Gaussian
distribution.
In recent empirical studies of 
stock market indices it was examined whether
the distribution 
$P(r)$
of returns 
$r(\tau)$ after some time $\tau$
can be described by a
(truncated) L\'evy-stable distribution 
$L_{\alpha}( r )$
with some index $0 < \alpha \le 2$.
While the L\'evy distribution cannot be expressed in
a closed form,
one can identify its parameters by
testing the dependence of the central peak 
height
on
$\tau$
as well as the power-law decay of the tails.
In an earlier study 
[Mantegna and Stanley, Nature {\bf 376}, 46 (1995)]
it was found that the behavior of the central peak
of $P(r)$ for the Standard \& Poor 500 index
is consistent with the L\'evy distribution
with $\alpha=1.4$. 
In a more recent study 
[Gopikrishnan et al., Phys. Rev. E {\bf 60}, 5305 (1999)]
it was found that 
the tails 
of $P(r)$
exhibit a power-law decay with
an exponent $\alpha \cong 3$,
thus deviating from the L\'evy distribution.
In this paper we study 
the 
distribution of returns
in 
a generic model 
that describes the dynamics of stock market indices.
For the distributions 
$P(r)$
generated by this model, we observe
that the scaling of the central peak is consistent
with a L\'evy distribution 
while the tails
exhibit 
a power-law distribution
with an exponent $\alpha > 2$, 
namely beyond the range of L\'evy-stable distributions.
Our results are in agreement with both empirical studies
and reconcile the apparent disagreement between 
their results.
\end{abstract} 

\pacs{PACS: 05.40.Fb,05.70.Ln,02.50.-r}

\newpage

\section{Introduction}

Financial time series are generated by 
complex dynamical processes that exhibit strong correlations
between many degrees of freedom.
The efforts to understand the dynamics of economic systems
have involved empirical studies 
in which
the temporal fluctuations 
of the prices of individual companies as well as
of stock market indices such 
as the Standard \& Poor 500 (S\&P500)
were examined
\cite{Mantegna1995,Gopikrishnan1999,Mantegna1998,Mantegna1999,Plerou1999,Gopikrishnan1998,Liu1999,Plerou2000}.   
These fluctuations can be characterized by the distribution
of stock market returns as well as the volatility,
that quantifies the magnitude of the market fluctuations. 

Consider a stock market index 
$\bar W(t)$.
Its value is proportional to the average of the
market values $W_i$, $i=1,\dots,N$ 
(given by the stock price of each firm times the number of its outstanding shares)
of the $N$ stocks that are included in this index.
The fluctuations of $\bar W$ can be expressed in terms of the returns
after a period of time $\tau$ (say, in minutes), given by

\begin{equation}
r(\tau) =  {\rm ln} \bar W(t+\tau) - {\rm ln} \bar W(t).
\label{eq:returnW}
\end{equation}

\noindent
For any value of $\tau$ one can examine the distribution 
$P(r)$
of the returns 
$r(\tau)$.
The number of independent data points available in the distribution is
given by $T/\tau$, where $T$ is the time period covered in the
available 
data set.
It was observed long ago that such distributions exhibit 
slowly decaying tails, unlike the Gaussian or exponential 
distributions. 
Moreover, the shape of the distribution was found to exhibit 
a self-similar form for different choices of $\tau$.
It was proposed 
by Mandelbrot
\cite{Mandelbrot1963}
that $P(r)$
may be expressed by a L\'evy-stable distribution,
$L_{\alpha}(r)$, where $0 < \alpha \le 2$
\cite{Levy1937,Gnedenko1954}.
Mathematically, the L\'evy distribution 
$L_{\alpha}(r)$ 
is the limit 
$n \rightarrow \infty$
of the
distribution of the sum of 
$n$ independent stochastic variables taken 
from a power-law distribution of the form 
$p(r) \sim r^{-1-\alpha}$ when
$0 < \alpha \le 2$
(that clearly exhibits an infinite variance).  
This is unlike the case of a distribution with a finite variance,
that leads to a Gaussian distribution of the sum, 
according to the central limit theorem.
The L\'evy distribution 
thus exhibits an infinite variance.
However, in practical applications its tail is truncated due
to an upper cutoff in the power-law distribution that generated
it
\cite{Mantegna1994b}.
Although the L\'evy distribution 
cannot be expressed in a closed form
\cite{Mantegna1994a},
it has two scaling properties that can be used in order to examine
whether a distribution 
$P(r)$
obtained from empirical data or numerical simulations
is a (truncated) L\'evy distribution and to calculate its index $0< \alpha \le 2$.
The first property involves the dependence of the central peak height
on the time $\tau$, that takes the form
\cite{Mantegna1994a}

\begin{equation}
L_{\alpha}(r=0) \sim \tau^{-1/\alpha}.
\label{eq:peak}
\end{equation}

\noindent
Thus, if the distribution of returns $P(r)$
is a (truncated) L\'evy distribution, 
the value of $\alpha$
can be obtained from the slope of the
graph of $P(r=0)$ vs. $\tau$
on a log-log scale.
The second property involves the power-law decay of the
tails of the distribution that
follows
\cite{Mantegna1994a}

\begin{equation}
L_{\alpha}(r) \sim r^{-1 - \alpha}.
\label{eq:tail}
\end{equation}

\noindent
Therefore, if the distribution $P(r)$
is a (truncated) L\'evy distribution, 
the value of $\alpha$
can also be obtained from the slope of the
tail of $P(r)$ vs. $r$
on a log-log scale.
Obviously, a L\'evy 
distribution should
satisfy the scaling relations for both the central peak and the
tail, with the same exponent 
$\alpha$.

The distribution $P(r)$ of the returns
$r(\tau)$
for the S\&P 500 stock market index 
was recently studied
for a range of $\tau$ values,
using the data for the six-year period of 1984-89
\cite{Mantegna1995}.
The scaling of the central peak height
vs. $\tau$
was examined 
within the range of
$1 \le \tau \le 1000$ minutes.
yielding a straight line in the log-log 
scale
over three orders of magnitude,
with a slope that corresponds to $\alpha=1.4$.
It was thus concluded
that 
$P(r)$
takes the form of 
a truncated L\'evy distribution
$L_{\alpha}(r)$ 
with the index $\alpha = 1.4$.
More recently the data set was extended to cover a 13-year period
(1984-96)
and
was examined using the scaling analysis of
the tail
of the distribution $P(r)$ of the returns $r(\tau)$
for $\tau$ in the range between 1 minute and 4 days.
\cite{Gopikrishnan1999}.
It was found that the tail of $P(r)$ vs. $r$, on 
a log-log scale exhibits a straight line domain,
indicating a power-law dependence
given by
Eq.~(\ref{eq:tail}).
However, the slope was found to be consistent with 
$\alpha$
in the range
$2.5 < \alpha < 3.5$,
where the precise value depends on details 
such as the value of $\tau$ and the fitting procedure.
Clearly, these values of $\alpha$ are
well outside the L\'evy-stable range of
$0 < \alpha \le 2$.
Therefore, not only that the distribution $P(r)$
is not a 
L\'evy 
distribution with
$\alpha=1.4$ - it is not a 
L\'evy 
distribution at all.
Apparently, this result seems to be in disagreement with the conclusions of
Ref. 
\cite{Mantegna1995}.
We thus observe that while the central peak maintains its L\'evy features
the tails show a non-L\'evy behavior.
In order to understand these puzzling results 
one needs to combine 
theoretical studies,
suitable models and simulations
of stock market dynamics, 
complementary to the empirical analysis.

In this paper we study the distribution of the returns $P(r)$
in a dynamical model that describes the time evolution of stock market
indices
\cite{Levy1996b,Solomon1996,Biham1998,Malcai1999}. 
The model consists of 
dynamic variables 
$w_i$, $i=1,\dots,N$ 
that represent the capitalization (total market values) 
of $N$ firms. 
The dynamics 
represents the increase (or decrease) by a random factor 
$\lambda(t)$ 
[taken from a predefined distribution $\Pi(\lambda)$]
of the value
$w_i$ of the firm $i$ between times $t$ and $t+1$. 
The dynamical rules also enforce a lower bound on the
$w_i$'s,
which is a certain fraction $0 \le c < 1$
of the momentary average of the $w_i$'s.
This lower bound
may represent
the minimal requirements for a company stock to be publicly traded.
It turns out that after some equilibration time the
$w_i$'s exhibit a power-law
distribution of the form
$p(w) \sim w^{-1-\alpha}$
\cite{Malcai1999}.
For any given value of $N$,
the exponent $\alpha > 0$ is a monotonically
increasing function of $c$.
Since $r(\tau)$ can be considered as a sum of $\tau$
random variables taken from a power-law  distribution $p(w)$,
one may expect it to converge to the 
L\'evy distribution $L_{\alpha}(r)$
with the same exponent $\alpha$.
Since the power-law distribution is truncated from above, the
tails of the resulting L\'evy distribution is also expected to
be truncated
\cite{Mantegna1994b}.
Clearly, the dynamics is much more complicated.
One reason for this is that the $\tau$ random variables are not
completely independent - they are taken from a finite set of
N values of the $w_i$'s. Moreover, these values slowly change
during the calculation of $r(\tau)$,
because at each time step one of the $w_i$'s is updated. 

To analyze the distribution of returns $P(r)$ 
we first tune the parameter $c$ 
(for the given value of $N$)
to adjust the power-law distribution
to the economically relevant case of $\alpha=1.4$
\cite{Mantegna1995,Levy1997}.
We then examine the distribution of returns $P(r)$
for a range of time intervals $\tau$
and test the scaling behavior of the central peak as well as of the
tails.
It is found that
the scaling of the central peak is consistent with
a truncated L\'evy distribution with $\alpha=1.4$
for a broad range of 
$ 1 \le \tau \le 1000$. 
For small values of $\tau$,
up to about $\tau=50$
(for $N=1000$)
the power-law decay of the tail of $P(r)$
is also consistent
with a truncated L\'evy distribution with 
the same value of $\alpha$.
However, 
for larger values of $\tau$
the tail of 
$P(r)$ exhibits a power-law decay
consistent with $\alpha > 2$,
and thus
deviates from the L\'evy
distribution.
These results are in agreement with the empirical analysis of
the central peak presented in Ref.
\cite{Mantegna1995}
as well as with the more recent analysis
of the tails
presented in Ref.
\cite{Gopikrishnan1999}.
They thus 
reconcile the apparent disagreement between these two empirical 
studies.

The paper is organized as follows. In Sec. II we present 
the model. Simulations and results are reported in Sec. III, followed by a 
summary in Sec. IV.

\section{The Model}

The model
\cite{Levy1996b,Solomon1996,Malcai1999}
describes the evolution in discrete time 
of $N$ dynamic variables $w_i (t)$, 
$i=1,\dots,N$. 
At each time step $t$, 
an integer $i$ is chosen randomly in the range
$1 \leq i \leq N$, 
which is the index of the dynamic variable $w_i$
to be updated at
that time step. 
A random multiplicative factor 
$\lambda(t)$ is then drawn from a
given distribution 
$\Pi(\lambda)$, which is independent 
of $i$ and $t$ and satisfies
$\int_{\lambda} \Pi(\lambda) d{\lambda} = 1$.
This can be, for example, a uniform 
distribution in the range
$\lambda_{min} \leq \lambda \leq \lambda_{max}$,
where
$\lambda_{min}$
and
$\lambda_{max}$
are predefined limits. 
The system is then updated 
according to the following stochastic time evolution equation

\begin{eqnarray}
w_i (t+1) &=&   \lambda (t)  w_i (t)  \nonumber \\w_j (t+1) &=& w_j (t), 
\ \ \ \ \ \ j=1,\dots,N; \ j \ne i.
\label{eq:mult}
\end{eqnarray}

\noindent
This is an 
asynchronous update mechanism. 
The average value of the system components at time t 
is given by

\begin{equation}
\bar w(t) = {1 \over N} \sum_{i=1}^N w_i(t).
\end{equation}

\noindent
The term on the right hand side of 
Eq.~(\ref{eq:mult}) 
describes the effect of 
auto-catalysis at the individual level. 
In addition to the update rule of 
Eq.~(\ref{eq:mult}),
the value of 
the updated variable 
$w_i (t+1)$ is constrained to be larger or equal to 
some lower bound which is proportional to 
the momentary average value of the $w_i$'s
according to

\begin{equation}
w_i (t+1) \ge c \cdot \bar w(t)
\end{equation}

\noindent
where $0 \le c < 1$ is a constant factor. 
This constraint is imposed immediately after step
(\ref{eq:mult})
by setting

\begin{equation}
w_i (t+1) \rightarrow \max \{ w_i(t+1), c \cdot \bar w(t) \},
 \ \ \ 
\label{eq:subsidize}
\end{equation}

\noindent
where 
$\bar w(t)$,
evaluated just before the application of 
Eq.~(\ref{eq:mult}),
is used. 
This constraint describes the effect of 
auto-catalysis at the community level.
Numerical simulations 
of the stochastic multiplicative process 
described by Eqs.
(\ref{eq:mult})
and
(\ref{eq:subsidize}), 
show that the $w_i$'s follow
a power-law distribution of 
the form

\begin{equation}
p(w) = K w^{-1-\alpha}
\label{eq:power}
\end{equation}

\noindent
for a wide range of lower bounds $c$,
where K is a normalization factor
\cite{Malcai1999}. 
It was found that the exponent
$\alpha$ depends on the parameters $c$ and $N$ 
and is insensitive to the 
shape of the probability distribution
$\Pi(\lambda)$. 
For simplicity,
we use
$\lambda$ 
uniformly distributed in the range
$0.9 \leq \lambda \leq 1.1$.

\section{Simulations and Results}
\label{simulations}

In the simulations below
the number of dynamical variables 
is $N=1000$ and the lower cutoff is chosen as $c=0.3$,
the value that provides 
the economically relevant distribution characterized by
$\alpha=1.4$
\cite{Mantegna1995,Levy1997}.
Under these conditions $p(w)$ exhibits
a power law distribution 
within three 
decades, between $w_{min}=0.0003$
and $w_{max}=0.3$.
The data for this distribution was obtained from a large
number of simulations  
collecting data at
different times within each simulation
after some equilibration time.
To remove the possible effect of inflation,
the values of the $w_i$'s fed into the distribution $p(w)$
were normalized such that at any time $t$
the sum $\sum_i w_i(t) =1$, namely
$\bar w(t) = 1/N$
\cite{Malcai1999}.
In the analysis of the returns, there is no need
for such normalization adjustment, due to the fact that
the returns quantify changes relative to the current value
of $\bar w$,
namely they are normalized by definition.

Consider the time evolution of the average $\bar w(t)$.
At each time step, when  
Eq.~(\ref{eq:mult}) 
is applied,
neglecting the effect of the lower cutoff we obtain

\begin{equation} 
\bar w(t+1)=\bar w(t) + {1 \over N} [\lambda(t)-1] w_i(t). 
\label{eq:rw}
\end{equation}

\noindent
This can be considered as 
a generalized random walk with step sizes 
distributed according to 
Eq.~(\ref{eq:power}).
Therefore, the returns
after $\tau$ time steps, given by

\begin{equation}
r(\tau) = {\rm ln} \bar w(t+\tau) - {\rm ln} \bar w(t)
\label{eq:return}
\end{equation}

\noindent
are expected 
\cite{Mantegna1994b} 
to follow a
a truncated L\'evy distribution $L_{\alpha}(r)$.
Note that for small time intervals, the returns
given by 
(\ref{eq:return})
coincide with the relative change given by

\begin{equation}
\tilde r(\tau) = {{{\bar w(t+\tau)} - {\bar w(t)}} \over
{\bar w(t)}}.
\label{eq:returnshort}
\end{equation}

\noindent
However, for large $\tau$ these two expressions provide significantly
different results.

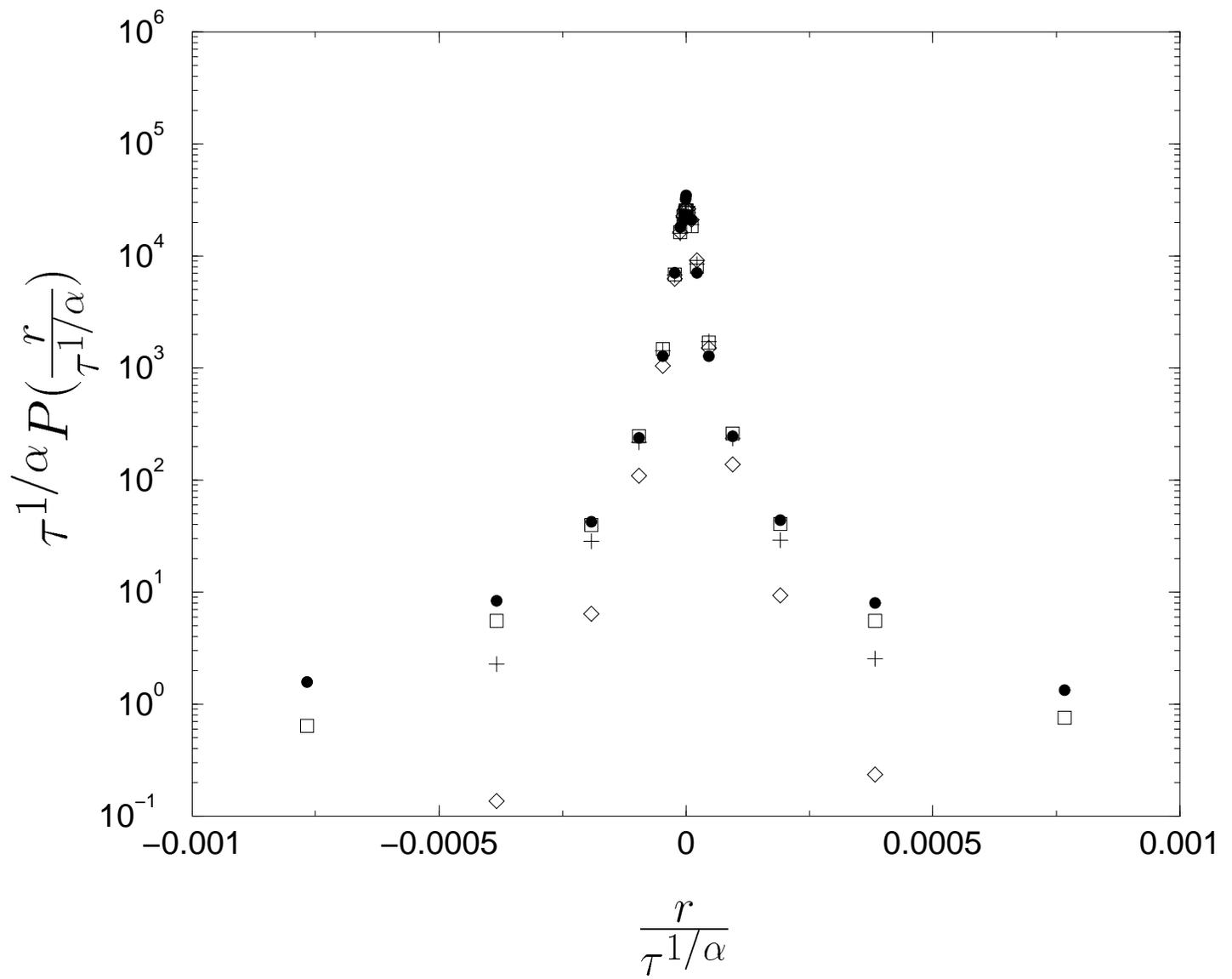
\begin{figure}
\caption{
The rescaled distribution of the returns 
$\tau^{1/\alpha} P(r(\tau)/\tau^{1/\alpha})$
for
$\tau=1$, 
50,
200
and 1000.
In the vicinity of the central peak we observe a  
collapse of all four graphs into a similar shape.
The tails for the two smaller values of $\tau$ follow
the  L\'evy-stable distribution with 
$\alpha =1.4$.
The tails for the two larger values of $\tau$ 
fall off more sharply and exhibit
significant deviations from the 
L\'evy-stable shape.
}
\label{fig:levy}
\end{figure}

In  
Fig.~\ref{fig:levy}
we show the 
rescaled
distribution
$\tau^{1/\alpha} P(r/\tau^{1/\alpha})$
of the returns 
$r(\tau)$ 
for $\tau = 1$, 50, 200 and 1000.
Near the central peak the four rescaled graphs 
collapse into a similar shape.
The graphs for $\tau=1$ and 50 maintain a similar
rescaled form also in the tails
while for larger values of
$\tau$ the tails go down more sharply.

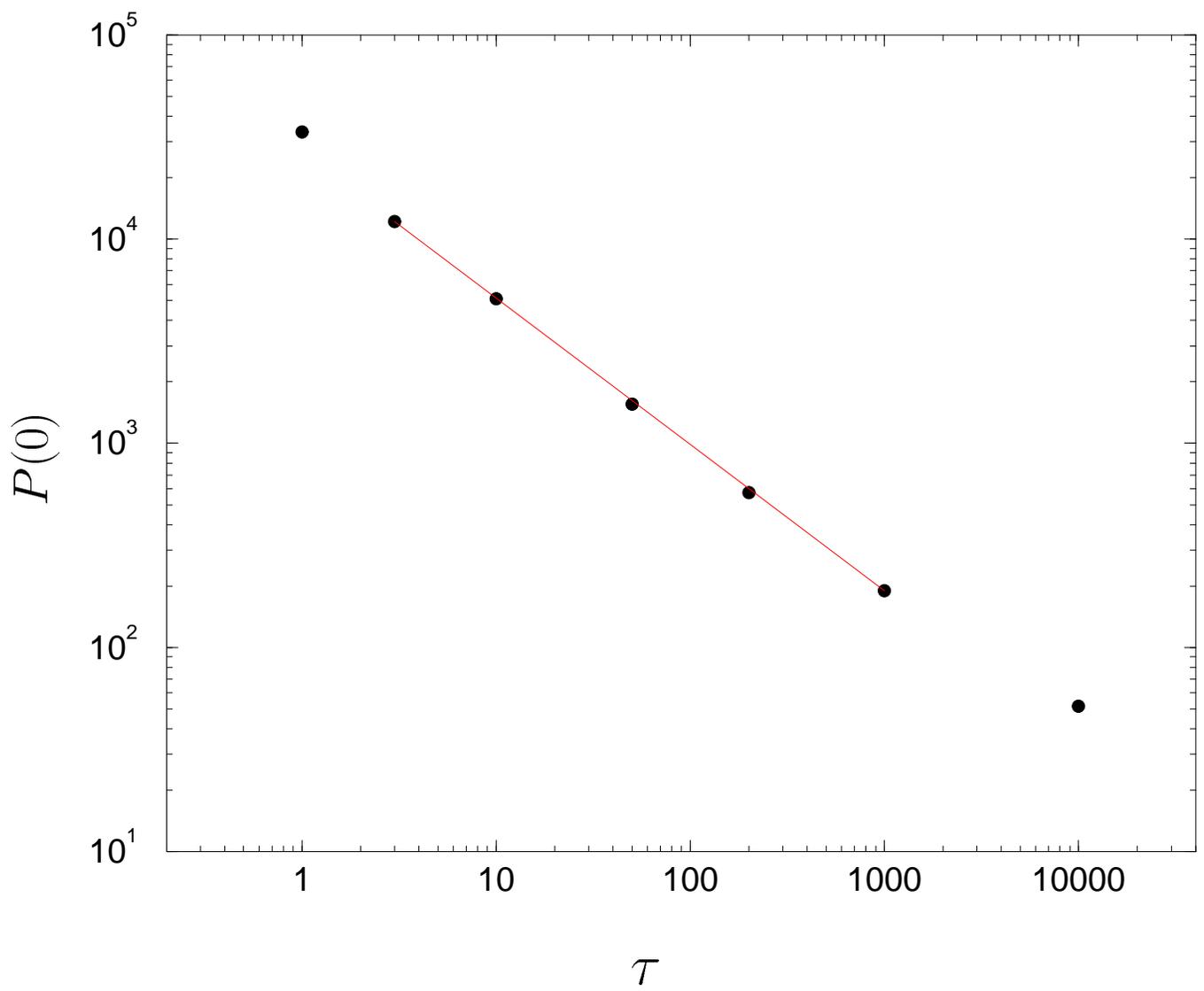
\begin{figure}
\caption{
The height of
the central peak 
$P(r(\tau) = 0)$ 
vs. $\tau$ on a log-log scale.
For a broad range of nearly three orders of magnitude in $\tau$
values up to $\tau=1000$,
the slope of the straight line is
$-1/\alpha=-0.71$,
which corresponds to a L\'evy distribution
with $\alpha = 1.4$.
}
\label{fig:levyzero}
\end{figure}

The value of $\alpha$
that characterizes the
distribution 
can be
obtained from the scaling of the central peak height
as a function of $\tau$,
according to 
Eq.~(\ref{eq:peak}).
In 
Fig.~\ref{fig:levyzero}
we show the height of the peak  
$P(r=0)$ 
as a function of  
$\tau$ on a log-log scale.
It is found that the slope of the fit 
is $-0.71$, which 
following  the scaling relation 
of
Eq.~(\ref{eq:peak}) 
means that the index of the
L\'evy distribution 
is $\alpha= -1/(-0.71) = 1.4$.

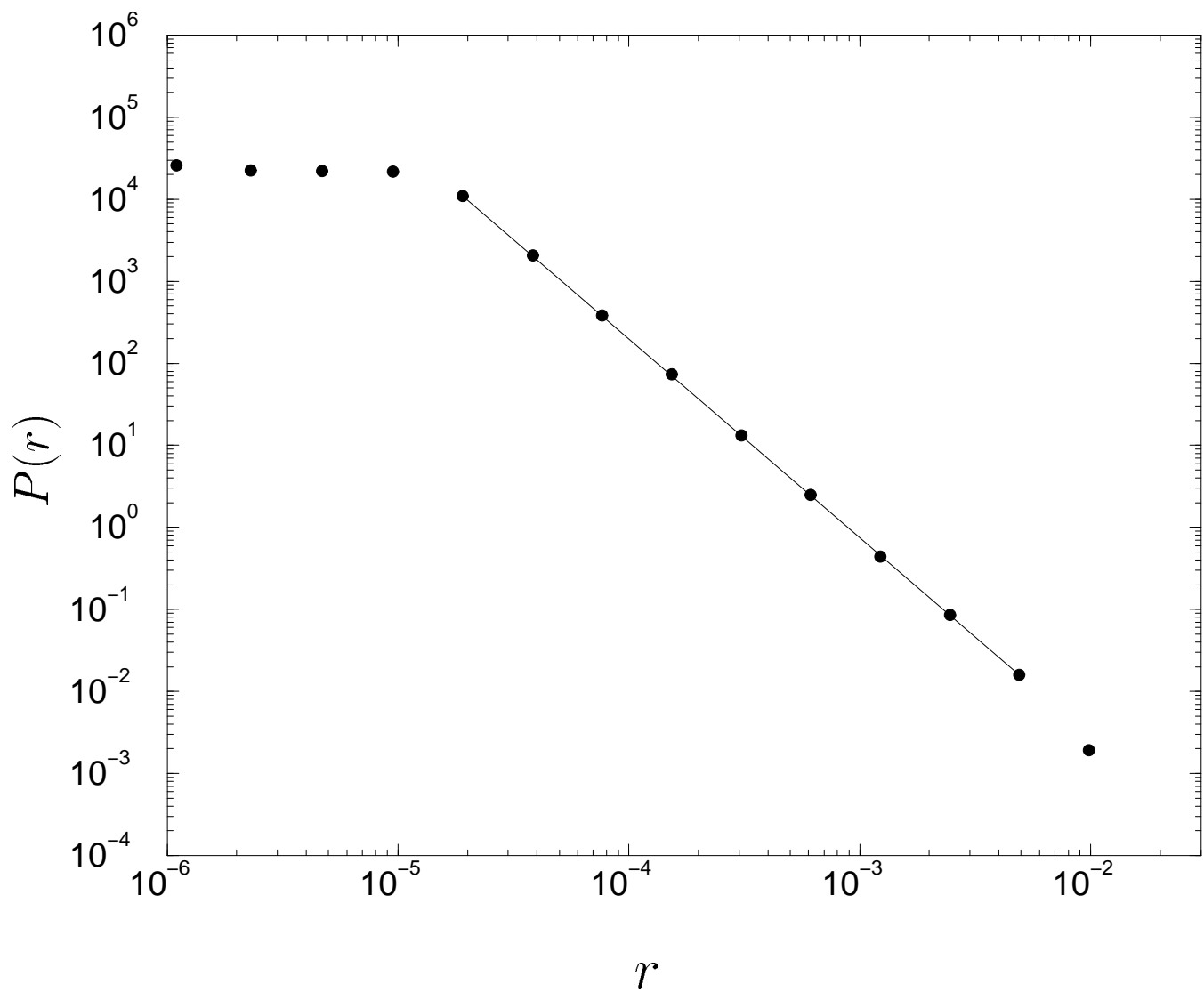
\begin{figure}
\caption{
The distribution 
$P(r)$
of $r(\tau)$
on a log-log scale,
for
$\tau=1$.
The tail exhibits a range of power-law behavior
according to 
Eq. (\ref{eq:tail})
with 
$\alpha =1.4$, 
namely following a L\'evy
distribution with the same value of $\alpha$.
}
\label{fig:levytail}
\end{figure}

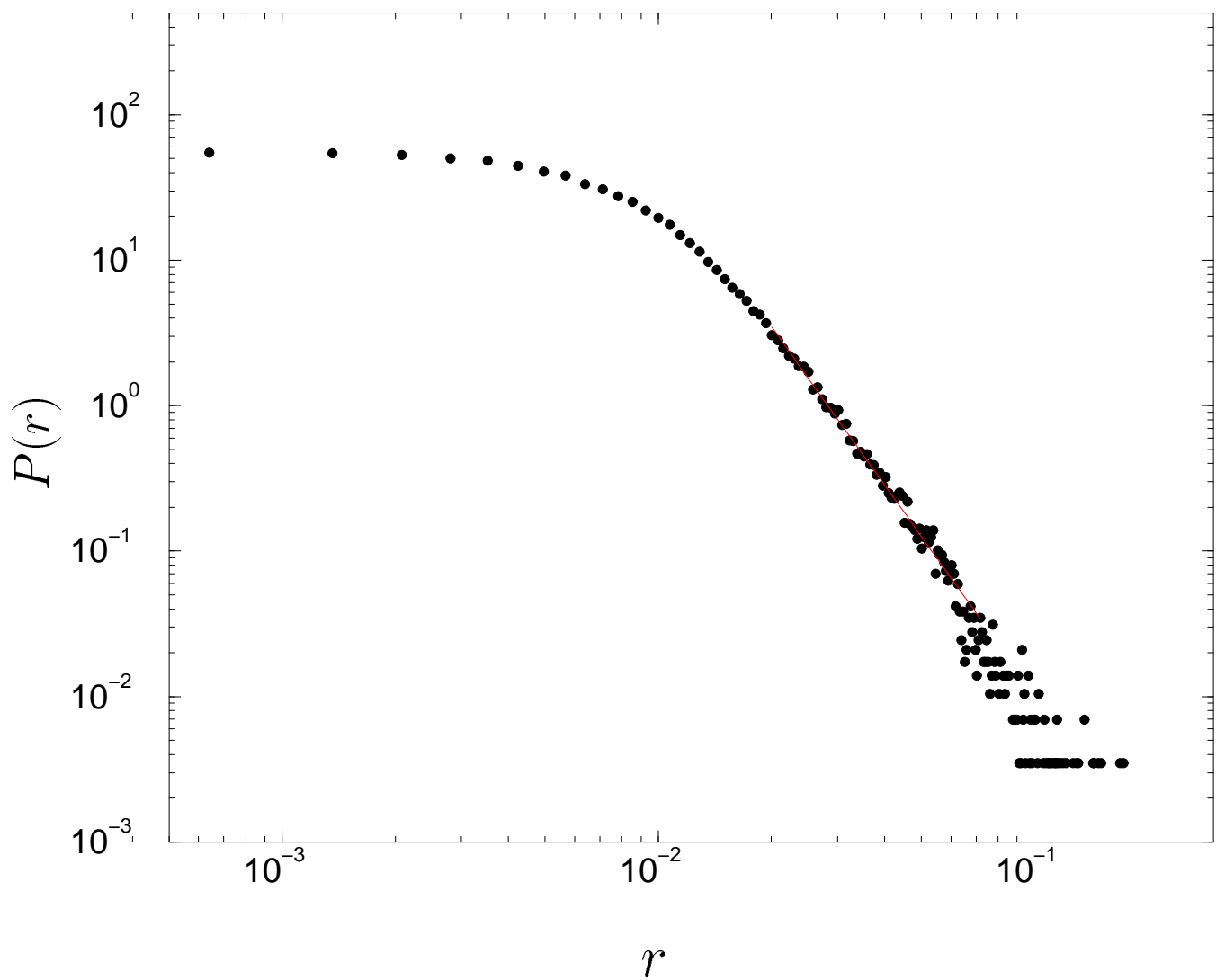
\begin{figure}
\caption{
The distribution 
$P(r)$
of $r(\tau)$
on a log-log scale,
for
$\tau=10^4$.
The tail exhibits a range of about one order of magnitude with
an apparent 
power-law behavior.
The slope in this range is consistent with
Eq. (\ref{eq:tail})
with 
$\alpha = 2.5$.
This value is not only 
different from the $\alpha=1.4$ observed
for short times, but is outside 
the range for L\'evy-stable distributions. 
This curve strongly resembles the empirical
distributions for the S\&P500 presented
in Ref.
[2].
}
\label{fig:levylongtail}
\end{figure}

To characterize the nature of the distribution
$P(r)$ we also examine the scaling behavior of the
tails. 
For the L\'evy distribution the tail is expected to
follow a power-law behavior 
given by Eq.~(\ref{eq:tail}).
In Fig.~\ref{fig:levytail}
we present the tail of the distribution $P(r)$,
on a log-log scale for $\tau=1$.
It is found that the slope is
$-(1+\alpha) =-2.4$ 
which corresponds to a L\'evy
distribution with $\alpha=1.4$.
For larger values of $\tau$, the tails exhibit
steeper slopes that exceed the domain
of the L\'evy distribution, 
namely $\alpha$ becomes larger than 2.
As an example, we present in 
Fig.~\ref{fig:levylongtail}
the distribution 
$P(r)$
of $r(\tau)$
for $\tau=10^4$
on a log-log scale.
We identify a range of about one order of magnitude in which the
apparent slope is
$-(1+\alpha)=-3.5$,
namely corresponds to $\alpha=2.5$,
which is outside the domain of the
L\'evy distribution.
It is thus observed that the tails of the distribution
$P(r)$ are much more sensitive to deviations from 
a L\'evy-stable process than the central peak.

These results 
are in agreement with the empirical analysis of
the central peak presented in Ref.
\cite{Mantegna1995}
as well as with the analysis
of the tails
presented in Ref.
\cite{Gopikrishnan1999}.
They thus reconcile the apparent disagreement between these two empirical 
studies.
To relate the parameters of the model more closely with the empirical
studies we note that the typical time required for a single stock-market
transaction is of the order of one minute.
However, the transactions are done simultaneously in all the stocks included
in the index that is analyzed.
Therefore, the single transaction-time unit (say, one minute) roughly corresponds,
in the model, to $\tau=N$ time steps.
The results of 
Fig.~\ref{fig:levylongtail}
for $\tau=10^4$
are thus expected to correspond to a time interval of several
minutes in the empirical analysis.
Indeed, the value of $\alpha=2.5$ obtained in the numerical 
simulations is only slightly lower than the empirical results
obtained for $\tau$ in the range between 1 and 512 minutes.

In the model we observe 
significant deviations from the 
L\'evy distribution as $\tau$ increases towards the order of $N$.
A possible explanation is that
at this stage 
some of the $w_i$'s are already sampled more than once in a sequence of $\tau$
time steps required to calculate one instance of $r(\tau)$.
This violates the requirement 
in the construction of a (truncated) L\'evy-stable distribution,
that the $\tau$ random variables should be independent.
This starts to introduce significant correlations between the
different variables that compose $r(\tau)$.

Another correlation effect is intrinsic to the calculation
of the returns. Consider the return $r(\tau)$, which is
given by

\begin{equation} 
r(\tau) = \sum_{t=1}^{\tau} ln \left[1 + [\lambda(t)-1]  {w_i(t) \over \bar w(t)} \right].
\label{eq:roftau}
\end{equation} 

\noindent
where the variable $w_i(t)$ 
is independently picked at any time t. 
Note that the return depends on the 
normalized quantities $w_i^{\prime} = w_i(t)/\bar w(t)$. 
It is easy to see that the $w_i^{\prime}$'s are not independent
since at any time $t$ they satisfy
$\sum_i w_i^{\prime} = N$.
This dependence is particularly apparent for the large $w_i^{\prime}$'s,
since if one of them turns out to be extremely large
the normalization condition prevents other $w_j^{\prime}$'s
from having values in its vicinity.

\section{Summary}

Recent empirical studies of the fluctuations in stock market
indices have provided conflicting results.
In these studies the 
distribution $P(r)$ of stock market returns $r(\tau)$ after time $\tau$
were examined.
The scaling of the central peak of $P(r)$
was found to be consistent with a (truncated) L\'evy-stable
distribution
with index $\alpha=1.4$
\cite{Mantegna1995}.
However, the scaling of the tails, 
for a broad range of $\tau$ values between 1 minute and a few days,
was found to exhibit a power-law
behavior with an exponent $\alpha \cong 3$, which is well outside the
range of the L\'evy distribution
\cite{Gopikrishnan1999}.

In this paper we have examined the distribution $P(r)$
for a model that describes the dynamics of
stock market indices.
The model consists of dynamical variables 
$w_i$, $i=1,\dots,N$,
that describe the
time-dependent market values of N firms, while their
average is the corresponding stock market index.
It was found that the scaling of the central
peak is consistent with a L\'evy distribution
and its index can be tuned to the economically relevant value
of $\alpha=1.4$ by tuning a parameter.
The tails of the distributions $P(r)$ of the returns $r(\tau)$, 
for a range of $\tau$ values that corresponds to the empirically
studied time intervals,
were found to exhibit a domain of power-law behavior 
with $\alpha > 2$,
that falls outside the range of the L\'evy distribution.
These results are fully consistent with the empirical results
both for the central peak and for the tails
and reconciles the apparent disagreement between them.

\end{document}